\documentclass[12pt]{article}
\usepackage{amsxtra,amssymb,amsthm,amsmath,latexsym}

\theoremstyle{plain}

\newtheorem{theorem}{Theorem}[section]
\numberwithin{equation}{section}

\def\R{{\mathbb R}}

\def\oH{{\overset{\circ}{H}}}
\def\oH1{{\overset{\circ}{H}\kern-.02in{}^1}}

\def\bee{\begin{equation*}}
\def\eee{\end{equation*}}
\def\be{\begin{equation}}
\def\ee{\end{equation}}

\begin{document}
American Mathematical Monthly, 120:8, (2013), 743-746.

%\begin{titlepage}
\title{ A variational principle and its application
to estimating the electrical capacitance of a perfect conductor}

\author{
A.G. Ramm}
%http://www.math.ksu.edu/\,$\widetilde{\ }$\,ramm}

\date{}

\maketitle\thispagestyle{empty}

\begin{abstract}
Assume that $A$ is a bounded
self-adjoint operator in a Hilbert space $H$. Then, the variational
principle
\begin{equation}\tag{*}\label{eq:*} \max_{v}\frac{|(Au,v)|^2}{(Av, v)}
= (Au, u) \end{equation}
holds if and only if $A \geq 0$, that is, if $(Av, v)
\geq 0$ for all $v \in H$.
 We define the quotient on the left-hand side in (*) to be zero if $(Av, 
v)=0$.
As an application of this principle a variational principle 
for the electrical capacitance of a conductor of an arbitrary shape is 
derived. 
%it is
%proved that \begin{equation}\tag{**}\label{eq:**} C = \max_{v \in
%L^2(S)}\frac{|\int_{S}vdt|^2}{\int_{S}\int_{S}\frac{v (t)v
%(s)dsdt}{4\pi |s - t|}}
\end{abstract}

%\footnote{2000 Math subject classification: 58C15, 47J06}
%\footnote{Key words:  nonlinear
%equations, homeomorphism, surjectivity, dynamical systems
%method (DSM)}

%\end{titlepage}

\section{INTRODUCTION.}\label{S:1}
In many applications, a physical quantity of interest can be
expressed as a quadratic form. For example, let $\sigma (t)$ be the
surface density of an electric  charge distributed on the surface $S$
of a perfect conductor, with $t$ being a point on $S$. If the
conductor is charged to a potential $u= 1$, then the equation for
$\sigma (t)$ is (see, for example, \cite{R144})
\begin{equation}\label{eq:1} A\sigma(s) := \int_{S}\frac{\sigma 
(t)dt}{4\pi
r_{st}} = 1,\quad s \in S,\quad r_{st} := |s - t|, \end{equation}
where $dt$ is the element of the surface area, $S$ is the surface of
the conductor $D$, and $D \subset\mathbb{R}^3$ is a bounded domain
with a connected smooth boundary $S$. The total charge on $S$ is $Q
= \displaystyle\int_{S}\sigma (t)dt$. The physical quantity of
interest is the electrical capacitance, $C$, of the  conductor $D$. Since
$Q = Cu$ and $u = 1$ (see equation (\ref{eq:1})), it follows that
$$C =\int_{S}\sigma (t)dt = (A\sigma, \sigma),$$ where $(f,g) :=
\displaystyle\int_{S}f\overline{g}dt$ is the inner product in the
Hilbert space $H:= L^2(S)$, and the overbar stands for complex
conjugation. The electrical capacitance of a perfect conductor of an
arbitrary shape is of interest from both the physical and mathematical
points of view. Our aim in this paper is to derive an abstract
variational principle that allows the representation of the
quadratic form $(Au,u)$ of a self-adjoint operator $A$ in a Hilbert
space $H$. If $A\ge 0$, then the variational principle (1.2) (see
below) follows from the Cauchy inequality. If $A$ is not
non-negative, it is not clear whether  the variational principle
(1.2) is true. In this paper, we prove that as long as this principle
holds, then $A$ must be non-negative. Our main result of this paper
is stated in Theorem 1.1. One physical application of this result is
the estimation of the electrical capacitance (see formulas (1.3) and
(1.4)). Several earlier works (see \cite{R476},\cite{R190},
\cite{R1}) provide various estimates of the electrical capacitances of
perfect conductors. In \cite{R2}, the role of the electrical
capacitance in the theory of wave scattering by small bodies of an
arbitrary shape is explained.

Let us introduce a general theory. Let $A = A^{\star}$ be a linear
self-adjoint bounded operator in a Hilbert space $H$. In the
abstract theory developed below, $u$ stands for an element of the
Hilbert space
$H$, while $\sigma$ is used in an application to electrostatics.

We are interested in  the quantity $(Au, u)$ and want to find a
variational principle that allows one to calculate or  estimate this
quantity. Let us write $A \geq 0$ if and only if $(Av, v) \geq 0$
for all $v$, and say in this case that $A$ is non-negative. If $(Av,
v) > 0$ for all $v \neq 0$, we write $A > 0$ and say that $A$ is
positive. The following variational principle is our main abstract
result.
\begin{theorem}\label{Theorem1} Let $A=A^{\star}$ be a linear
bounded self-adjoint operator. The formula
\begin{equation}\label{eq:2}
(Au, u) = \max_{v \in H}\frac{|(Av, u)|^2}{(Av, v)}
\end{equation}
holds if and only if $A \geq 0$.
\end{theorem}

{\bf Remark.} If $(Av, v)=0$, then the  quotient on the right-hand side 
of equation
(\ref{eq:2}) is defined to be zero.

Theorem 1.1 can be proved also for unbounded self-adjoint operators
$A$. In that case, maximization is taken over $v\in D(A)$, where
$D(A)$ is the domain of $A$, a linear dense subset of $H$, and 
it is assumed that $u\in D(A)$.

In Section 2, Theorem \ref{Theorem1} is proved. Let us illustrate this
theorem with an example.

$\\$ {\bf Example of an application of the variational principle
(1.2).} Let $A$ be defined as in equation (\ref{eq:1}). In Section 2, we
prove the following lemma.

%\begin{lemma}\label{Lemma1.3}
{\bf Lemma 1.2.} {\it The operator $A$ in equation (\ref{eq:1}) is
positive in $H := L^2(S)$.}
%\end{lemma}

From Theorem \ref{Theorem1}, Lemma 1.2, and equation (\ref{eq:1}), it
follows that the electrical capacitance $C$ can be calculated by the
following variational principle:
\begin{equation}\label{eq:3}
C = \max_{v \in L^2(S)}\frac{|\int_{S}v(t)dt|^2}{\int_{S}\int_{S}
\frac{v(t)\overline{v(s)}dsdt}{4\pi r_{st}}}.
\end{equation}

This variational principle for electrical capacitance is an example
of the application of the abstract variational principle formulated
in Theorem 1.1.

Formula (\ref{eq:3}) can be rewritten as
\begin{equation}\label{eq:4}
C^{-1} = \min_{v \in
L^2(S)}\frac{\int_{S}\int_{S}\frac{v(t)\overline{v(s)}
dsdt}{4\pi r_{st}}}{|\int_{S}v(t)dt|^2}.
\end{equation}

In particular, by setting $v = 1$ in (\ref{eq:3}) one gets
\begin{equation}\label{eq:5}
C \geq \frac{4\pi|S|^2}{J},\quad and \quad J :=
\int_{S}\int_{S}\frac{dsdt}{r_{st}},
\end{equation}
where $|S|$ is the surface area of $S$.

Formula (\ref{eq:4}) yields a well-known principle, due to Gauss (see
\cite{[PS]}). This principle states that if the total charge $Q =
\displaystyle\int_{S}v(t)dt$ is distributed on the surface $S$ of a
perfect conductor with a density $v(t)$, then the minimal value of
the functional
\begin{equation}\label{eq:5b}
Q^{-2}\int_{S}\int_{S}\frac{v(t)\displaystyle\overline{v(s)}dsdt}
{4\pi r_{st}} = \min
\end{equation}
is equal to $C^{-1}$. Here $C$ is the electrical capacitance of the
conductor, and this minimal value is attained at $v(t) = \sigma(t)$,
where $\sigma (t)$ solves equation (\ref{eq:1}).

In \cite{R476}, the following approximate formula for the
capacitance is derived:
$$C^{(0)} = \displaystyle\frac{4\pi|S|^2}{J}.$$
This formula is the zero-th
approximation of an iterative process for finding $\sigma (t)$, the
equilibrium charge distribution on the surface $S$ of a perfect
conductor
charged to the potential $u = 1$.

\section{PROOFS}\label{S:2}
%\label{Proofs}

\begin{proof}[Proof of Theorem \ref{Theorem1}] The {\it sufficiency} of
the
condition $A \geq 0$ for the validity of (\ref{eq:2}) is clear: If $A =
A^{*} \geq 0$, then the quadratic form $[u,u] := (Au, u)$ is
non-negative
and the standard argument yields the Cauchy inequality
\begin{equation}\label{eq:6} |(Au, v)|^2 \leq (Au, u)(Av, v).
\end{equation}
The equality sign in formula (\ref{eq:6}) is attained if and only
if $u$ and $v$ are linearly dependent. Dividing (\ref{eq:6}) by
$(Av, v)$, one obtains (\ref{eq:2}), and the maximum in (\ref{eq:2}) is
attained if $v = \lambda u$ and $\lambda$ is constant.

Let us prove the {\it necessity} of the condition $A \geq 0$ for
(\ref{eq:2})
to hold. Let us assume that there exist $z$ and $w$ such that $(Az, z) >
0$
and $(Aw, w) < 0$, and prove that then formula (\ref{eq:2}) cannot hold.

Note that if $(Av, v) \leq 0$ for all $v$, then (\ref{eq:2}) cannot
hold.
Indeed, if $(Av, v) \leq 0$ for all $v$, then (\ref{eq:2})
%implies $|(Au,v)|^2 \geq (Au, u)(Av, v)$. This inequality
can be written as $|(Bu, v)|^2\geq (Bu, u)(Bv, v)$, where $B = -A \geq 0$.
This is a contradiction to
the Cauchy inequality. This contradiction proves that
inequality $(Av, v) \leq 0$ cannot hold for
all $v$ if the variational principle  (\ref{eq:2}) is true.

Continuing with our proof, take $v = \lambda z + w$, where $\lambda$
is an arbitrary real number. Then, (\ref{eq:2}) yields
\begin{equation}\label{eq:7}
\frac{|(Au, \lambda z + w)|^2}{q(\lambda)} \leq (Au, u),
\end{equation}
where
\begin{equation}\label{eq:8}
q(\lambda) := a\lambda^2 + 2b\lambda + c,
\quad a := (Az, z) > 0,\quad c
=(Aw, w) < 0,
\end{equation}
and $b := Re(Az, w)$. The polynomial $q(\lambda)$ has two real zeros,
$\lambda_1 < 0$ and $\lambda_2 > 0$, and $q^{-1}(\lambda) \to +\infty$
if $\lambda \to \lambda_1 - 0$ or if $\lambda \to \lambda_2 + 0$.
The quadratic polynomial $p(\lambda) := |(Au, \lambda z + w)|^2$ also has
two zeros, and by (\ref{eq:7}), the ratio
$\frac{p(\lambda)}{q(\lambda)}$ is bounded when $\lambda \to
\lambda_1 - 0$ and $\lambda \to \lambda_2 + 0$. Therefore, one
concludes that $p(\lambda)$ has the same zeros as $q(\lambda)$, that
is, $\lambda_1$ and $\lambda_2$ are zeros of $p(\lambda)$.

Since $\lambda_1\lambda_2 < 0$ and
$$p(\lambda) = |(Au, z)|^2\lambda^2 + 2\lambda \text{Re}(Au, z)
\overline{(Au, w)} + |(Au, w)|^2,$$ it follows that
\begin{equation}\label{eq:9}
\frac{|(Au, w)|^2}{|(Au, z)|^2} < 0.
\end{equation}
This is a contradiction which proves that there are no elements $z$ and
$w$ such that $(Az, z) > 0$ and $(Aw, w) < 0$.
Theorem \ref{Theorem1} is proved.
\end{proof}

\begin{proof}[Proof of Lemma 1.2]
The idea behind the proof of Lemma 1.2 is simple. The Fourier
transform of a quadratic form $(Au,u)$,  with an operator $A$ that
has a convolution kernel, is non-negative  for every $u$ if and only
if the Fourier transform of the convolution kernel $A(x)$ (that is the
function $\tilde{A}(\zeta)$) is non-negative. In our case, the role
of $u$ plays the density $\sigma$ supported on the smooth surface
$S$, that is, $u=\sigma \delta_S$, where $\delta_S$ is the 
delta-function supported on the surface $S$, and 
the convolution kernel is $A(x)=\frac 1{|x|}$ in $\R^3$.

 It is known  that
\begin{equation}\label{eq:10}
F(\frac{1}{|x|}):=\int_{\mathbb{R}^3}\frac{e^{-i\zeta \cdot x}}{|x|}dx =
\frac{4\pi}{|\zeta|^2} > 0,
\end{equation}
where the Fourier transform $F$ is understood in the sense of
distributions
(see, e.g., \cite{[GS]}). Therefore,
\begin{equation}\label{eq:11}
(A\sigma, \sigma) = \int_{S}\int_{S}\frac{\sigma (t)\overline{\sigma
(s)}}
{4\pi|s - t|}dsdt = \int_{\mathbb{R}^3}\frac{|F\sigma(\zeta)|^2}
{|\zeta|^2}d\zeta \geq 0,
\end{equation}
where  $F\sigma(\zeta)$ is the Fourier transform of the distribution
$\sigma(t)$ with support on the surface $S$. There are many results
regarding the rate of decay of the Fourier transform of a function
(measure) supported on a surface. For example, if the Gaussian
curvature of the surface $S$ is strictly positive, then (see
\cite{[S]})
\begin{equation}
F\sigma(\zeta): = \int_{S}\sigma (t)e^{-i\zeta \cdot t}dt =
O\bigg(\frac{1}{|\zeta|}\bigg),\quad |\zeta| \to \infty, 
%\quad \zeta \in \mathbb{R}^3,
\end{equation}
where $ |\zeta| \to \infty$, $ \zeta \in \mathbb{R}^3$, and we assume 
that $\sigma(t)$ is sufficiently smooth. With such a decay
the integral in  (\ref{eq:11}) converges classically. It follows
from  (\ref{eq:11}) that $(A\sigma, \sigma)\ge 0$. This proves Lemma
1.2.
\end{proof}

%\newpage


\begin{thebibliography}{1000}
%number of characters oflongest bibitem label

\bibitem{[GS]} I. Gel'fand, G. Shilov, {\it Generalized functions.} 
Acad. Press, New York, 1964.

\bibitem{[PS]} G. Polya, G. Szeg\"{o}, {\it Isoperimetric inequalities
in
mathematical physics.} Princeton University Press, Princeton, 1951.


\bibitem{R144} A. G. Ramm, {\it Iterative methods for calculating
static fields
 and wave scattering by small bodies.} Springer-
 Verlag, New York, 1982. 

\bibitem{R476} A. G. Ramm, {\it Wave scattering by small bodies of
arbitrary shapes.} World
Scientific Publishing, Singapore, 2005.

\bibitem{R190} A. G. Ramm, {\it  Scattering by obstacles.} D.Reidel,
Dordrecht, 1986.

\bibitem{R1} A. G. Ramm, A variational principle and its applications, 
{\it Internat. Journ. of
Pure Appl. Math.} {\bf 77} no. 3 (2012) 309-313.

\bibitem{R2} A. G. Ramm, Scattering of scalar waves by many
small particles, {\it AIP Advances} {\bf 1} (2011) 022135.

\bibitem{[S]} E. Stein, {\it Harmonic analysis: real-variable methods,
orthogonality and oscillatory integrals.} Princeton University Press,
Princeton, 1993.
%\end{thebibliography}
$$
$$











%Alexander G.Ramm
Department of Mathematics, Kansas State University, Manhattan, KS 
66506-2602,


email:     ramm@math.ksu.edu






\end{thebibliography}
\end{document}